\documentclass[iop]{emulateapj}
\usepackage{apjfonts}

\usepackage{amsmath,amssymb}

\newcommand{\myemail}{minoways@subaru.naoj.org}
\newcommand{\lya}{Ly$\alpha$}
\newcommand{\ha}{H$\alpha$}
\newcommand{\mgii}{Mg~{\sc ii}}
\newcommand{\siii}{Si~{\sc ii}}
\newcommand{\hi}{H~{\sc i}}
\newcommand{\oiii}{[O~{\sc iii}]}

\slugcomment{Accepted for publication in The Astronomical Journal}

\shorttitle{Constraining stellar properties of intervening DLA and
  \mgii\ absorbers toward GRB 050730}
\shortauthors{Minowa et al.}

\begin{document}

\title{CONSTRAINING STELLAR PROPERTIES OF INTERVENING DAMPED \lya\ and
  \mgii\  ABSORBING GALAXIES TOWARD GRB 050730}

\author{Y. Minowa\altaffilmark{1}, K. Okoshi\altaffilmark{2},
  N. Kobayashi\altaffilmark{3}, and H. Takami\altaffilmark{4}}

\altaffiltext{1}{Subaru Telescope, National Astronomical Observatory
  of Japan, 650 North A'ohoku Place, Hilo, HI 96720, USA \myemail} 
\altaffiltext{2}{Faculty of Industrial Science and Technology, Tokyo
  University of Science, 102-1 Tomino, Oshamambe, Hokkaido, 049-3514, Japan}
\altaffiltext{3}{Institute of Astronomy, School of Science, University
  of Tokyo, 2-21-1 Osawa, Mitaka, Tokyo 181-0015, Japan}
\altaffiltext{4}{Optical and Infrared Astronomy Division, National
  Astronomical Observatory of Japan, 2-21-1 Osawa, Mitaka, Tokyo
  181-8588, Japan}

\begin{abstract}
  We performed multi-band deep imaging of the field around GRB 050730 to
  identify the host galaxies of intervening absorbers, which
  consist of a damped \lya\ absorption (DLA) system at
  $z_{\rm{abs}}=3.564$, a sub-DLA system at $z_{\rm{abs}}=3.022$, and strong
  \mgii\ absorption systems at $z_{\rm{abs}}=1.773$ and $2.253$. Our
  observations were performed after the gamma-ray burst afterglow had disappeared.
  Thus, our imaging survey has a higher sensitivity to the host galaxies
  of the intervening absorbers than the normal
  imaging surveys in the direction of QSOs, for which the QSO glare tends to hide the
  foreground galaxies. In this deep imaging
  survey, we could not detect any unambiguous candidates for the host
  galaxies of the intervening absorbers. Using the 3-$\sigma$ upper
  limit of the flux in the optical to mid-infrared observing bands, which
  corresponds to the UV to optical bands in the rest-frame of the intervening
  absorbers, we constrained the star-formation rates and stellar
  masses of the hosts. We estimated the star-formation rates for the intervening
  absorbers as $\lesssim 2.5$ $M_{\odot}$ yr$^{-1}$ for
  $z>3$ DLAs and $\lesssim 1.0$ $M_{\odot}$ yr$^{-1}$ for $z\sim2$
  \mgii\  systems. Their stellar masses are estimated to be several times
  $10^9$ $M_{\odot}$ or smaller for all intervening galaxies. These
  properties are comparable to dwarf galaxies, rather than the
  massive star-forming galaxies commonly seen in the $z>2$ galaxy
  surveys based on emission-line selection or color selection.
\end{abstract}


\section{INTRODUCTION}
Exchange of gas and metals between the intergalactic medium and
galaxies is a fundamental process of galaxy formation and evolution.
In the standard model of galaxy formation based on the cold dark
matter (CDM) cosmology \citep[e.g.,][]{white78, fall80}, galaxies are
thought to be formed by star formation caused by the cooling and
condensation of gas in the core of galactic halos produced by the
hierarchical clustering of dark matter. To reveal the process of
galaxy formation and evolution, it is important to explore the gaseous
and stellar properties of young galaxies at high redshifts. The study
of stellar populations in high-redshift ($z>2$) galaxies has been
largely developed by the detection of color-selected galaxies, such as
Lyman break galaxies \citep[LBGs;][]{steidel03}, or emission-line
selected galaxies, such as \lya\ emitters
\citep[LAEs;][]{cowie98}. The study of these galaxy populations
provides us with the properties of their stellar population, including
star formation rate (SFR), stellar mass, and age. The gaseous properties of
high-redshift galaxies, such as hydrogen and metal abundances, have
also been studied in this decade, but these studies are mostly based
on observations of intergalactic matter identified as hydrogen
or metal absorption lines in the spectra of bright QSOs. It is
difficult to connect the stellar properties of the emission-line
selected galaxies to the gaseous properties of the absorption-line
selected galaxies, as they have very different selection biases. The
emission-selected samples are biased by their flux and depend on the
depth of the imaging surveys. Therefore, they only provide information
in galaxies at the brightest end of the luminosity function. In
contrast, the absorption-selected samples are selected homogeneously
by their gas content, and thus, they are less biased by the flux of
the host galaxies. To connect stellar and gaseous properties of
high-redshift galaxies, it is essential to detect the emission from
the galaxies associated with the high-redshift absorption line
systems.

Among the several absorption lines seen in the spectra of background
QSOs, the damped \lya\ absorption (DLA) systems \citep{wolfe86}
present unique opportunities to select \hi-rich (\hi\ column density
($N_{\rm{HI}}$) $\geq 10^{20.3}$ cm$^{-2}$) galaxies at high redshift. The
DLAs dominate the \hi\ content of the universe at $z>2$
\citep[e.g.,][]{peroux03b,prochaska09, noterdaeme09} and contain a
large amount of \hi\ gas mass, accounting for a large fraction of the
present-day stellar mass \citep{cole01}. Thus, the high-z galaxies
associated with absorbers such as DLAs are expected to be progenitors
of present-day galaxies. Many attempts to identify DLA host galaxies
have been made so far. At low redshift ($z\leq 1$), a few dozen DLA
absorbing galaxies have been established \citep[e.g.,][]{rao03,
  chen03, rao11}, although more than 50\% of the known low-redshift
DLAs remain unidentified. At high redshift ($z\geq1$), however, the
success rate for discovering DLA host galaxies is quite low,
especially at $z>2$. Only a few DLA host galaxies have been identified
\citep[e.g.,][]{moller02, moller04, peroux12}. This low discovery rate
could be caused by the fact that most DLA host galaxies are faint and
compact. The glare of the background QSOs also severely hampers the
detection of the faint stellar emission from the DLAs. Some
independent theoretical predictions suggest that DLA host galaxies
have typical luminosities of sub-$L_*$ with a typical proper size of
$\sim 3$ kpc, and 60\%-90\% of them could be hidden by the glare of
bright QSOs \citep[e.g.,][]{fynbo99, okoshi05}. The limited sample
size of identified galaxies associated with DLAs prevents us from
continuing investigations to connect the stellar and gaseous
properties of high-redshift galaxies. Recently, \citet{fynbo10} have
initiated an emission-line survey for metal-rich DLAs whose
luminosities are expected to be brighter than typical less metal-rich
DLAs \citep{moller04,ledoux06}. They successfully identified three DLA
host galaxies at $z>2$ in nebular emission \citep{fynbo10,fynbo11,
  noterdaeme12}. As for typical less metal-rich DLA host galaxies,
only a few have been successfully detected to date. Even in the cases
of the identified DLA host galaxies, the contamination from the QSO
light obscures the detailed study of stellar properties such as
luminosity and morphology. Some of the identified galaxies have been
identified only by nebular emission (such as \lya, \oiii, \ha). To
detect the stellar continuum from absorbing galaxies unambiguously and
to increase the number of identified hosts, we have conducted a deep
imaging survey for host galaxies associated with intervening absorbers
toward gamma ray burst (GRB) afterglows. Because the GRB afterglow
dims on a short timescale, observations in the direction of GRBs
several months after their emergence allow for more sensitive searches
for absorbing galaxies than do those in the direction of QSOs, for
which the contamination from background lights is significant. Many
optical spectroscopy observations of GRB afterglows in the hours after
their bursts have been performed to date. Some of these have shown
signs of intervening ($z_{\rm{abs}} < z_{\rm{GRB}}$) absorbers
\citep[e.g.,][]{metzger97, mirabal02, vreeswijk04, chen05}. Several
attempts have been made to identify the galaxies giving rise to the
intervening absorption systems in the direction of the GRB
afterglow. Some of these have successfully identified the galaxies
associated with DLAs or metal-absorption systems at $z\sim1$
\citep{vreeswijk03, pollack09}.

In this context, GRB 050730 is a very unique target that has
intervening DLA features at $z>2$, where the discovery rate for
identifying the DLA host galaxies toward QSOs is quite low. GRB 050730
was first detected by the $Swift$ satellite \citep{holland05} on 2005 July
30. \citet{chen05} performed high-resolution ($\sim$ 10 km s$^{-1}$)
echelle spectroscopy of GRB 050730 4 hr after the initial burst,
when the afterglow was bright at $R \sim 17.7$, and confirmed the
redshift of its host galaxy at $z_{\rm{GRB}}=3.969$ based on a strong DLA
feature ($\log N_{\rm{HI}} \sim 22.15$). Additionally, the authors also
identified four other intervening absorption systems \citep{pro07a}:
$z_{\rm{abs}}=3.564$ DLA system ($\log N$(\hi) $\sim$ 20.3),
$z_{\rm{abs}}=3.022$ sub-DLA system ($\log N$(\hi) $\sim$ 19.9), and
$z_{\rm{abs}}=1.773, 2.253$ strong \mgii\ absorption systems, whose
rest-frame equivalent widths ($W_r^{\lambda2796}$) were $\sim 1$
\AA. \citet{chen05} estimated the metallicities for the intervening
DLA and sub-DLA systems at $z=3.564$ and $3.022$ in terms of silicon
abundance using one of the unsaturated \siii\ absorption lines associated
with the each system. They found that both of the intervening DLA and
sub-DLA systems show low-metallicities with [Si/H] $< -1.3$ and $-1.5
\pm 0.2$, respectively, which are typical for $z\sim3$ DLAs
\citep[e.g.,][]{fynbo08}. Strong \mgii\ absorption systems
($W_r^{\lambda2796} > 0.3$ \AA) are often found to be associated with
DLA or Lyman-limit systems \citep[$\log N$(\hi) $>$ 17.7; e.g.,][]{churchill00, rao06} and are known to be a strong probe
of the intergalactic medium along the line of sight toward QSOs or
GRBs.  Therefore, deep imaging of the field around GRB 050730 is
particularly important for detecting or constraining the stellar
properties of the intervening DLAs or \mgii\ absorption systems at $z
>2$ for the first time.

In this paper, we present the results of optical to near-infrared deep
imaging observations of the field around GRB 050730. The layout of the
paper is as follows: the optical and near-infrared observations and
data analysis are presented in Section 2. In Section 3, we provide the results of
our spectral energy distribution fitting analysis and discuss the
properties of the intervening absorbers toward GRB 050730.  Our main
results are summarized in Section 4. In this paper, we adopt a $\Lambda$CDM
cosmology with $\Omega_{\Lambda} = 0.7$, $\Omega_{M} = 0.3$ and $H_{0}
= 70$ km s$^{-1}$ Mpc$^{-1}$.

\section{OBSERVATIONS AND RESULTS} 
\subsection{Optical and Near-infrared Imaging}
We observed the field around GRB 050730 in 2007 May and June using the
Multi-Object InfraRed Camera and Spectrograph
\citep[MOIRCS;][]{suzuki08} mounted on the Cassegrain focus of the
Subaru telescope at Mauna Kea, Hawaii. We took images in the $J$ and
$K_s$ bands, with total exposure times of 3600 and 11160 s,
respectively. Seeing sizes were roughly 0\farcs51 in the $J$ band and
0\farcs58 in the $K_s$ band. A standard star \citep[FS136;][]{leggett06} was observed for
flux calibration at the beginning or end of each observation. The data
were reduced using a purpose-built pipeline software package called
MCSRED \citep{tanaka11}. First, we performed flat fielding with
self-flat frames made from the actual science images, and then we
subtracted the sky background from each image. Finally, the data were
co-registered and combined.

We also obtained $B$-, $R_c$-, and $z^{\prime}$-band optical images
of the GRB 050730 field using SuprimeCam \citep{miyazaki02} mounted at
the prime focus of the Subaru telescope. The total exposure times for $B$,
$R_c$, and $z'$ bands were 1950, 4000, and 2530 s,
respectively. The seeing size varied between 1\farcs2 and 1\farcs5
throughout the night of the observations. The data were reduced with
standard procedures (i.e., dark and bias subtraction, flat fielding
and sky subtraction, distortion correction, point-spread function (PSF) matching, and
mosaicking) using the SDFRED package \citep{yagi02, ouchi04}. The images
were calibrated with observations of the SA104 standard stars
\citep{landolt09} at an airmass similar to the observations of the
GRB 050730 field. 

\begin{deluxetable*}{crrrrr}
\tabletypesize{\scriptsize}
\tablecaption{Summary of the Observations and Data.\label{tbl-1}}
\tablewidth{0pt}
\tablehead{
\colhead{Band} & \colhead{Instrument} & \colhead{Date} &
\colhead{Exp. Time (s)} & \colhead{Seeing ($^{\prime\prime}$)} &
\colhead{Limiting Magnitude\tablenotemark{a}}
}
\startdata
\multicolumn{6}{l}{Subaru imaging}\\
$B$ & SuprimeCam & 2007 Jun. 13 & 1950 & 1.6 & 26.21 \\
$V$ & FOCAS & 2010 May 5& 180 & 0.8 & 25.73 \\
$R_c$ & SuprimeCam & 2007 Jun. 13& 4150 & 1.5 & 26.48 \\
$z^{\prime}$ & SuprimeCam & 2007 Jun. 13 & 2530 &  1.5 & 25.10 \\
$J$ & MOIRCS & 2007 Jun. 8 & 3600 & 0.5 & 25.22 \\
$K_s$ & MOIRCS   & 2007 May 1& 11160 & 0.6 & 25.04 \\
\hline
\multicolumn{6}{l}{Imaging data from archive}\\
$R_{\rm{special}}$ & FORS2/VLT &  2006  Feb.$-$May & 9800 & 0.7 & 27.69 \\
$I$ & FORS2/VLT & 2006 Feb. 25& 600 & 1.0 &  24.96 \\
$L$(3.6$\mu$m) & IRAC/$Spitzer$ & 2008, Mar. 11. & 6970 & 1.9 & 24.65 \\
\hline
\multicolumn{6}{l}{Subaru spectroscopy}\\
300B$+$SY47 & FOCAS &  2010 May 5 & 5400  & 0.7 & \nodata \\
\enddata
\tablenotetext{a}{All magnitudes are shown as $3\sigma$ limiting
  magnitudes measured in 2 $\times$ FWHM apertures. The data were
  corrected for Galactic extinction \citep{schlegel98}. }
\end{deluxetable*}

\begin{deluxetable*}{crrrrrrrr}
\tabletypesize{\scriptsize}
\tablecaption{List of the Intervening Absorbers Toward GRB 050730\label{tbl-2}}
\tablewidth{0pt}
\tablehead{
\colhead{Type} & \colhead{Redshift} & \colhead{$\log N$(\hi)} &
\colhead{\mgii\ $W_{r}^{\lambda2796}$} & \colhead{[Si/H]} &
\colhead{$M_{\rm{AB}}$} & \colhead{$L/L_*$\tablenotemark{a}} & 
\colhead{SFR}  & \colhead{$\log (M_{\rm{stellar}}/M_{\odot})$} \\
\colhead{} & \colhead{} & \colhead{(cm$^{-2}$)} &
\colhead{(\AA)} & \colhead{} &
\colhead{} & \colhead{} & \colhead{($M_{\odot}$ yr$^{-1}$)}  & \colhead{}}
\startdata
DLA & 3.564 & 20.03$\pm$0.10 & \nodata &  $< -1.3$ 
& $>-20.79$ ($B$) & $< 0.22$
& $<1.62$\tablenotemark{b}, $2.45$\tablenotemark{c}
& $<9.52$\tablenotemark{b}, $<9.92$\tablenotemark{c}
\\
sub-DLA & 3.022 & 19.90$\pm$0.10 & \nodata & $-1.5 \pm 0.2$ 
& $>-20.49$ ($V$) & $< 0.12$
& $<1.24$\tablenotemark{b}, $1.87$\tablenotemark{c}
& $<9.68$\tablenotemark{b}, $< 9.92$\tablenotemark{c}
\\
\mgii & 2.253 & \nodata & 0.886$\pm$0.031 & \nodata
& $>-19.95$ ($R$) & $< 0.08$
& $<0.75$\tablenotemark{b}, $1.14$\tablenotemark{c}
& $<9.66$\tablenotemark{b}, $<9.68$\tablenotemark{c}
\\
\mgii & 1.773  & \nodata & 0.943$\pm$0.023  & \nodata
& $>-19.49$ ($I$) & $< 0.03$
& $<0.49$\tablenotemark{b}, $0.74$\tablenotemark{c}
& $<9.71$\tablenotemark{b}, $<9.54$\tablenotemark{c}
\enddata
\tablenotetext{a}{The $3\sigma$ upper limit in the $K_s$-band image
  (Table \ref{tbl-1}), which corresponds to the rest-frame optical at the
 redshift of the absorbers ($B$ for $z=3.564$, $V$ for $z=3.022$, $R$ for $z=2.253$, and $I$ for
  $z=1.773$), was used for calculating the upper limit of $L/L_*$. The
  $L_*$ magnitudes for $B$, $V$, and $R$ bands are drawn from a
  luminosity function at $2<z<3.5$ from \citet{marchesini07}, and the
  $L_*$ magnitude for the $I$ band is drawn from a luminosity function at
$z<2$ from \citet{ilbert05}.}
\tablenotetext{b}{The case that no counterpart associated with the
  absorbers is detected in the deepest $R$-band image ($R>27.14$).}
\tablenotetext{c}{The case that the candidate G1 ($R\sim26.69$) is
  associated with the absorber at each redshift.}
\end{deluxetable*}%
\
\subsection{Optical and Mid-infrared Archived Images}
The field around GRB 050730 was imaged with several telescopes to
detect its optical afterglow just after the burst
\citep[e.g.,][]{pandey06}. Moreover, subsequent follow-up deep
imaging surveys to detect its host galaxy were made more than several
months after the burst when the optical afterglow has disappeared
\citep{chen09}. We used the archived images to specify the position of
the afterglow of GRB 050730 and to provide additional leverage to
identify the candidate of the galaxy associated with intervening
absorbers.

We used the $R$-band snapshot images of the field around GRB 05073
observed on 2007 July 30 (just after the burst), with Very Large
Telescope (VLT)/FORS2,
obtained from the ESO archive. We also used the VLT/FORS2 $R$ and
$I$-band deep imaging data observed in 2008 February$-$May (more
than six months after the burst; i.e., no afterglow in the image). The
total exposure times of the deep imaging data are 9800 and 600 s
in the $R$ and $I$ band, respectively. A basic calibration including
bias subtraction, flat fielding, and sky subtraction was performed on
each image using the EsoRex FORS pipeline. The resultant images have
seeing sizes of 0\farcs7 and 1\farcs0 in the $R$ and $I$ band,
respectively.

We also obtained the $L$-band (3.6$\mu$m) $Spitzer$/IRAC images of
the field of GRB 050730 from the $Spitzer$ archive
\citep{laskar11}. The total exposure time was 6970 s. The
data were reduced using the standard Spitzer Science Center pipeline
software, whose outputs are Basic Calibrated Data (BCD) FITS files for
each frame. The image mosaics were created from the BCD data using
the MOPEX software. The FWHM of the PSF in the final image is
estimated to be around 1\farcs9. 

\begin{figure*}
  \epsscale{.90}
\plotone{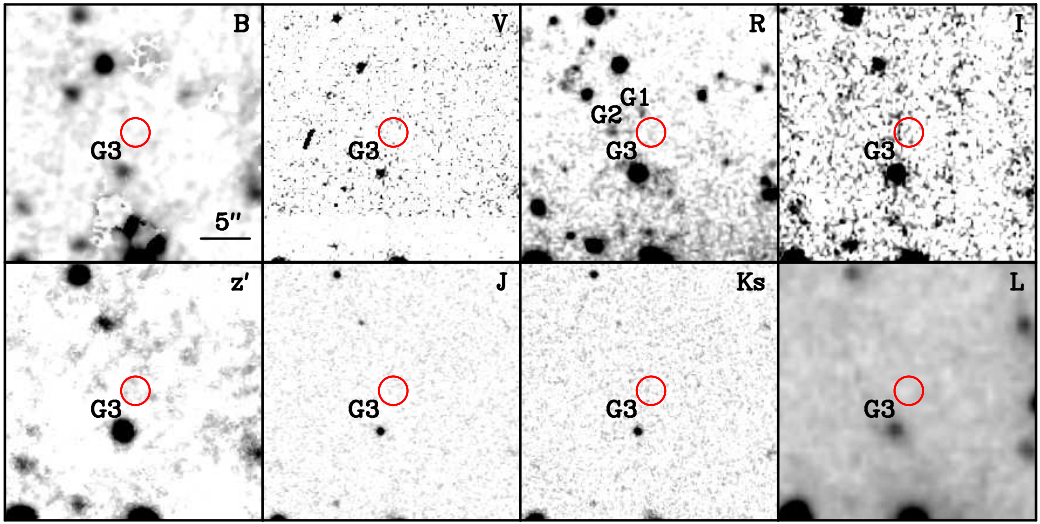}
\caption{$BVRIz^{\prime}JK_sL$(3.6$\mu$m)-band images of the field
  around GRB 050730. The size of each panel is 25$^{\prime\prime}$
  $\times$ 25$^{\prime\prime}$, which corresponds to $\sim$200 kpc
  $\times$ 200 kpc at the redshift of the intervening absorbers
  ($z_{\rm{abs}} \sim 1.5$-3.5). North is up, and east is to the
  left. A red circle in each panel shows the location of the optical
  afterglow of GRB 050730. The radius of the circle is around 10 kpc
  at the GRB's redshift ($z=3.969$), the area in which most GRB host
  galaxies were discovered \citep{bloom02, pro07b}. No detectable
  object is seen within 10 kpc around the GRB afterglow, as reported
  previously in the literature \citep{chen09}. \label{montage}}
\end{figure*}

\subsection{Limiting Magnitude\label{limit_mag}}
To estimate the limiting magnitude, we derived the $1\sigma$
background fluctuation in each band by directly measuring the sky
fluxes within 2 $\times$ FWHM apertures randomly
placed on the images. Table \ref{tbl-1} summarizes the $3\sigma$
limiting magnitude within a 2 $\times$ FWHM diameter aperture and
seeing size in each band, as well as the observational information. The limiting magnitudes listed in this table are
corrected for Galactic extinction toward the field around GRB 050730
\citep[$A_V=0.156$;][]{schlegel98}. Hereafter, we used the magnitude
corrected for Galactic extinction.
 
The $R$-band image of VLT/FORS2 is the deepest among the obtained
images. The limiting magnitude of the $R$-band image is 27.69 (AB),
which is more than 1 mag deeper than the rest. Therefore, we
use the $R$-band image for detecting galaxies around GRB 050730.
However, we do not use the $R$-band image for deriving the luminosity
because the $R$-band corresponds to the rest-frame UV wavelength at
the absorbers' redshift ($z>1.7$), and thus mainly traces the
star formation activities in galaxies rather than the stellar
properties. The $K_s$ band corresponds to the
rest-frame optical band at the absorbers' redshift ($B$ for $z_{\rm{abs}}=3.969$
and $3.564$, $V$ for $z_{\rm{abs}}=3.022$, $R$ for $z_{\rm{abs}}=2.253$, and $I$ for
$z_{\rm{abs}}=1.773$), and thus traces the long-lived stellar population in
the galaxies. Therefore, we used the $K_s$-band image to obtain the galaxy
luminosity or its upper limit. Note that although the $L$
(3.6$\mu$m)-band image also traces the rest-frame optical wavelength
at the absorbers' redshift, its limiting magnitude ($24.65$) is brighter than
that of the $K_s$ band ($25.04$; see Table \ref{tbl-1}).

We derived the absolute magnitude and luminosity in the rest-frame
optical for each absorber's redshift corresponding to the $3\sigma$
limiting magnitude in the $K_s$ band (Table \ref{tbl-2}). The luminosity
is expressed in terms of $L_*$ by assuming the $L_*$ magnitude drawn
from the rest-frame optical luminosity function at $2<z<3.5$
\citep[][$M_*(B)=-22.39$ for $z_{\rm{abs}}=3.564$, $M_*(V) = -22.77$ for
$z_{\rm{abs}}=3.022$, $M_*(R) = -22.67 $ for $z_{\rm{abs}}=2.253$]{marchesini07}
and at $z<2$ \citep[][$M_*(I) = -23.15 $ for
$z_{\rm{abs}}=1.773$]{ilbert05}. We emphasize that the limiting magnitude
for our $K_s$-band image corresponds to a luminosity less than
0.3$L_*$ at any absorber's redshift, and we note that we can reach the
sensitivity necessary to detect sub-$L_*$ galaxies, which are expected to be major contributors to the DLA cross-section \citep{okoshi05}.

\subsection{Sources around GRB 050730}
Figure \ref{montage} shows the acquired images of the field around the
GRB 050730 afterglow. The size of each panel is 25$^{\prime\prime}$
$\times$ 25$^{\prime\prime}$, which roughly covers the area around
$\sim$100 kpc from the line of sight toward the GRB afterglow at the
absorbers' redshift ($z_{\rm{abs}} \sim 1.7-3.9$). We detected 27 galaxies
within this field of view in the deepest $R$-band image.  Only seven
out of the 27 galaxies are detected in the $K_s$ band.

Long-duration GRBs are known to occur exclusively within a few
kiloparsecs of the center of star-forming galaxies \citep{bloom02, fruchter06}. We confirmed that there was no discernible flux
associated with the intrinsic GRB$-$DLA absorbers at $z \sim 3.968$
within 10 kpc around the afterglow, as already reported in
\citet{chen09}, although our limiting magnitude in $R$ band ($\sim
27.4$) is 1 mag deeper than that of \citet[][$R_c\sim26.4$]{chen09}.

\begin{figure}
\epsscale{1.1}
\plotone{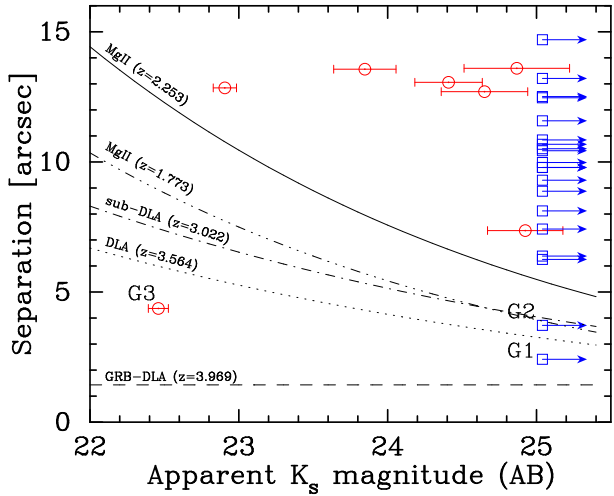}
\caption{Apparent $K_s$ magnitude vs. separation from the GRB
  afterglow for galaxies detected within the 25$^{\prime\prime}$
  $\times$ 25$^{\prime\prime}$ field-of-view around the afterglow. Circles
  show the galaxies detected in the $K_s$ band. Squares show
  the galaxies not detected in the $K_s$ band but detected in the $R$ band,
  and then indicate the $3\sigma$ upper limit of the $K_s$-band
  magnitude. The dashed line marks the upper limit of the impact
  parameter ($\sim 10$ kpc) for the host galaxies of GRB 050730 at
  $z=3.969$. The triple-dot-dashed line, solid line, dot-dashed line, and dotted line mark the upper limits of
  the impact parameter of the DLA or \mgii\ host galaxies at $z=1.773$,
  $2.253$, $3.022$, and $3.564$, respectively. We estimated the upper
  limits of the extension of cold gas clouds associated with DLA or
  \mgii\ host galaxies using an empirical relation between 
 the galaxy luminosities ($L$) and radius ($R$), which is
  described by a power law of the form $R= 43 (L/L_*)^{0.26}$ (kpc)
  for DLAs \citep{chen03} and $R = 107 (L/L_*)^{0.35}$ (kpc) for \mgii\ systems \citep{chen10}.
  \label{impact}}
\end{figure}

By contrast, intervening DLA or \mgii\ absorbers are selected
independently of any emission or stellar population. Their sight lines
are selected by a cross section of the intervening galaxies and should
preferentially intersect the outer regions of the interstellar medium in the galaxies.
To identify possible candidates for the galaxies associated with the
intervening absorbers, we first eliminate galaxies that are located
too far away from the line of sight toward the GRB afterglow to produce the
observed DLA or \mgii\ absorptions. We adopted a scaling relation between
the galaxy's luminosity ($L$) and halo radius ($R$) in the form
\begin{equation}
  \frac{R}{R_*} = \left(\frac{L}{L_*}\right)^{t}
\end{equation}
to constrain the extension of the gas cloud associated with the DLA or
\mgii\ host galaxies following the method described in \citet{fynbo99}. In this
formula, the halo radius $R_*$ and power-law index $t$ are determined
by their fit to the observed DLA or \mgii\ host galaxies at low redshift ($z
< 1$) as $R_*= 42.86$ kpc and $t=0.26$ for the DLA \citep{chen03}, and $R_*
= 107.14$ kpc and $t=0.35$ for the \mgii\ system \citep{chen10}. Based on
the semi-analytic models of galaxy evolution
\citep[e.g.,][]{mo99}, the halo size of galaxies is expected to be
smaller at higher redshift, which has also been confirmed in several
observational studies of disk galaxies
\citep[e.g.,][]{barden05,trujillo06}. Therefore, we used the radius
$R$ derived from the above formula as an upper limit of the extension
of the gas cloud associated with DLA or \mgii\ host galaxies at $z\geq
2.0$. Figure \ref{impact} shows the apparent separation between
the position of 27 detected galaxies in the $R$ band and the position of
the afterglow as a function of their apparent magnitude (or $3\sigma$
upper limit) in the $K_s$ band, which corresponds to the rest-frame optical
band at the redshift of the absorbers. We also plot the upper limits of the
halo radii associated with the DLA at $z=3.564$, sub-DLA at $z=3.022$,
and \mgii\ system at $z=2.253$ and $1.773$ based on the above
formula. The magnitudes corresponding to $L_*$ are drawn from the
rest-frame optical luminosity function as in Section
\ref{limit_mag}. We found that three galaxies are located below the
limit of the halo radius in Figure \ref{impact}. We identified these three
galaxies as candidates for the galaxies associated with the intervening
DLA or \mgii\ absorption systems toward GRB 050730. The three
candidates are marked G1, G2, and G3 in Figures \ref{montage} and
\ref{impact} in order of the apparent distance from the position of
the afterglow. G1 and G2 are faint and were only detected in the deepest
$R$-band image, whereas G3 is the brightest among the three candidates
and was detected in all bands.

\begin{figure*}
  \epsscale{0.7}
  \plotone{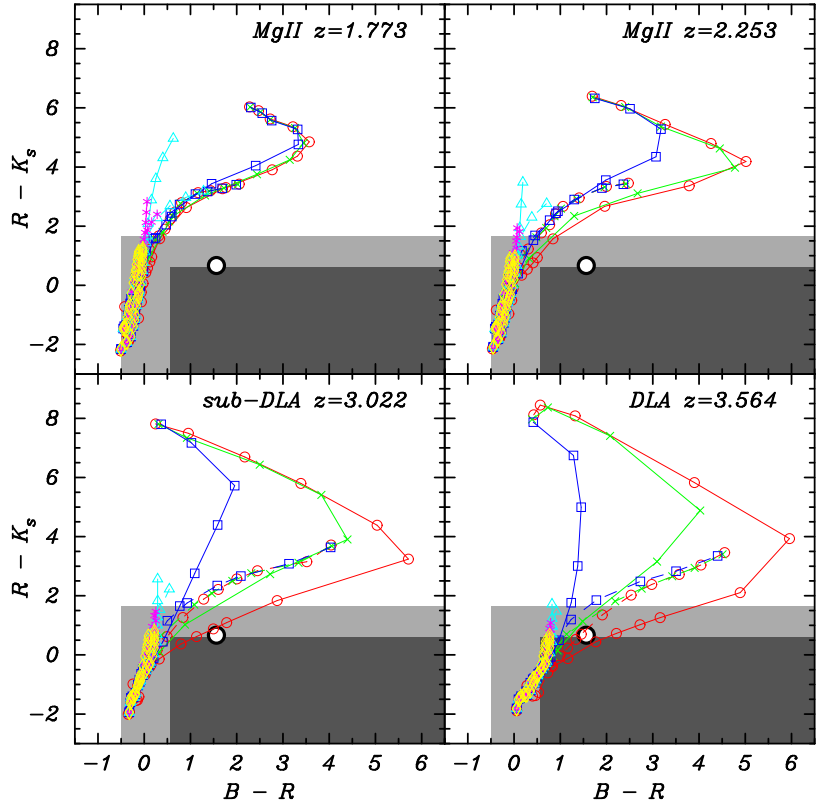}
  \caption{Evolutionary tracks in the two-color ($B-R$) vs. ($R-K_s$)
    diagrams at the redshift of the intervening absorbers from
    \citet{bc03} stellar population synthesis models. Stellar ages in
    the tracks range from 1 Myr to the cosmic age at each redshift.
    Single-stellar population (circles), exponentially decaying star
    formation with $e$-folding time $\tau$ of 0.05, 0.1, 0.5, and 1
    Gyr (crosses, boxes, triangles, and asterisks, respectively) and
    constant star formation (diamonds) models are used in these
    tracks. No dust attenuation is assumed. The solid and dashed lines
    show the evolutionary tracks for metallicities of $Z_{\odot}$ and
    1/50 $Z_{\odot}$, respectively. Thin and thick gray shaded areas
    show the range of color expected from the $3\sigma$ upper limit
    of $B$- and $K_s$-band magnitudes (see Table \ref{tbl-1}) and the
    $R$-band magnitude of the candidate G1 ($R\sim26.69$) and G2
    ($R\sim25.65$), respectively. The thick open circle shows the
    color of candidate 3.
\label{BR_RK}}
\end{figure*}%

\begin{figure*}
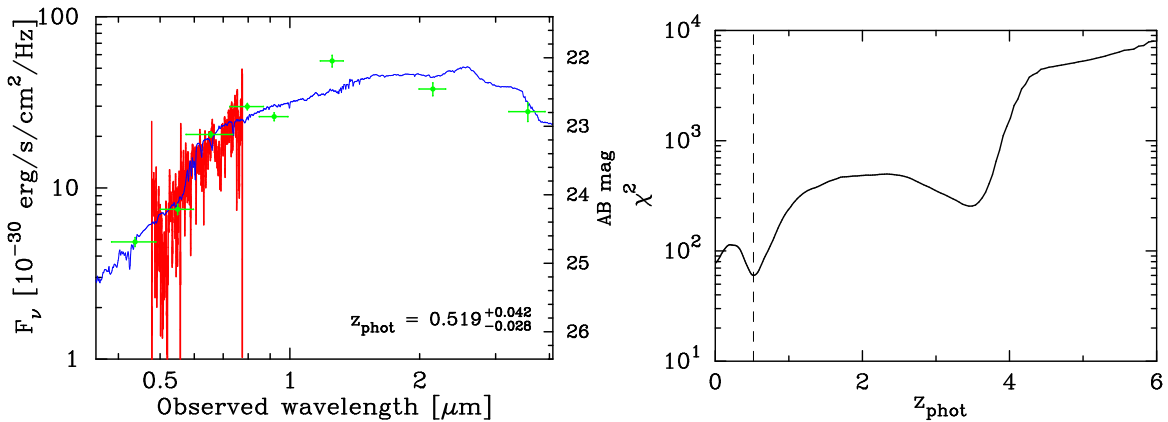

  \begin{center}
  \includegraphics[width=0.45\textwidth]{f4a.eps}
  \includegraphics[width=0.40\textwidth]{f4b.eps}
  \end{center}
 \caption{Left: photometric measurements of the candidate G3 (circles with error bars) and
    the best-fit SED (thin line) to the photometry data. The optical spectrum obtained by FOCAS is also
    plotted (thick line). The candidate G3 is most likely a low-redshift galaxy at $z\sim0.5$. Right: $\chi^2$ value of the SED
    fitting as a function of the photometric redshift.
    \label{phot-z}}
\end{figure*}

\subsection{Photometric Color Selection\label{color-color}}
To further constrain the candidates for the host galaxies of the
intervening absorbers, we used the stellar population synthesis models
from \citet{bc03} and examined a possible evolutionary track in the
two color ($B-R$) versus ($R-K_s$) diagrams at each absorber's
redshift (Figure \ref{BR_RK}). Stellar ages in the tracks range from 1
Myr to the cosmic age at each redshift. Single-stellar population,
exponentially decaying star formation with $e$-folding time $\tau$ of
0.05, 0.1, 0.5, and 1 Gyr, and constant star-formation models were
used in the evolutionary tracks. We also used two different
metallicities, $Z_{\odot}$ and 1/50 $Z_{\odot}$. We adopted the
\citet{salpeter55} initial mass function (IMF) with mass cutoffs of
0.1 and 100 $M_{\odot}$ for all cases. No dust attenuation was
assumed, as typical DLAs show extremely low dust attenuation with
$E(B-V) < 0.04$ \citep[e.g.,][]{ellison05, frank10}, and the
attenuation should be negligible in this two-color diagram. In Figure
\ref{BR_RK}, we also show the colors for the candidates, G1, G2, and
G3. For G1 and G2, because the exact $B$ and $K_s$ magnitudes were not
available, we include the possible ranges of color expected from the
$R$-band magnitude and the brightest limit of the $B$- and $K_s$-band
magnitudes (see Table \ref{tbl-1}). We found that the color range for
G1 (thin gray shaded area in Figure \ref{BR_RK}) overlaps with any
type of evolutionary track at any redshift. The color of G3 (circle in
Figure \ref{BR_RK}) can also be explained by the SED expected from the
single stellar population tracks at $z=3.022$ and $z=3.564$. These
indicate that G1 could be a candidate for the galaxy associated with
one of the four intervening absorbers, and G3 could be a candidate for
the galaxy associated with either $z=3.022$ (sub-DLA) or $z=3.564$
(DLA) intervening absorbers. The color range for G2 (thick gray-shaded
area in Figure \ref{BR_RK}) overlaps with the evolutionary tracks at
$z=3.564$ and marginally overlaps with single stellar population
tracks at $z=3.022$ at an age of around 40 Myr. This also indicates
that G2 could be a candidate for the galaxy associated with either of
the intervening absorbers at $z=3.022$ (sub-DLA) or $3.564$
(DLA). However, the impact parameter of G2 is larger than the size
limit for $z=3.564$ DLA and too close to the size limit for $z=3.022$
sub-DLA. Thus, G2 is unlikely to be a galaxy associated with the
intervening DLA or \mgii\ absorption systems toward GRB 050730.

\begin{figure*}
  \epsscale{0.88}
\plottwo {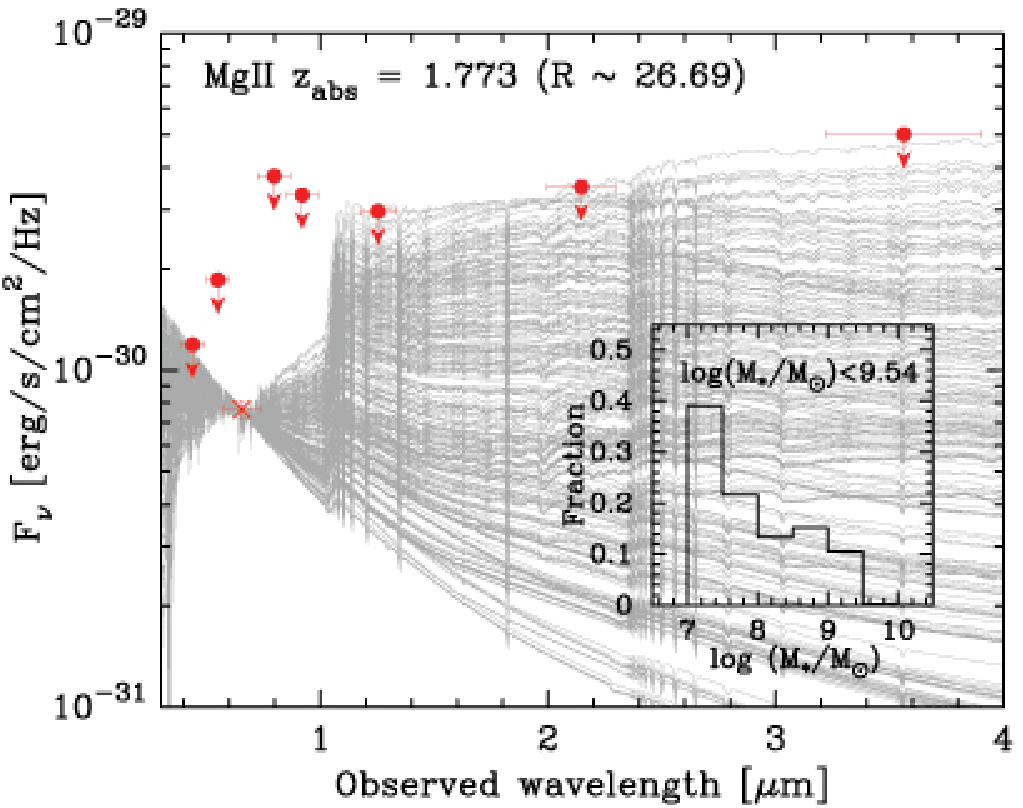}{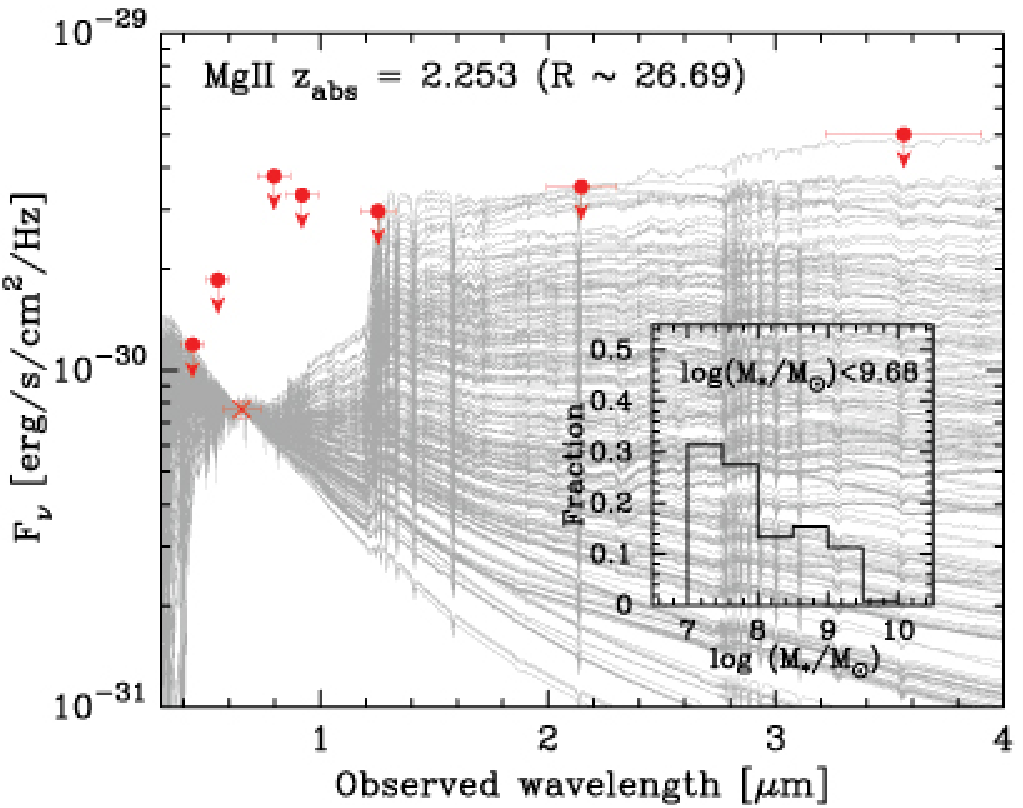}\\
\plottwo {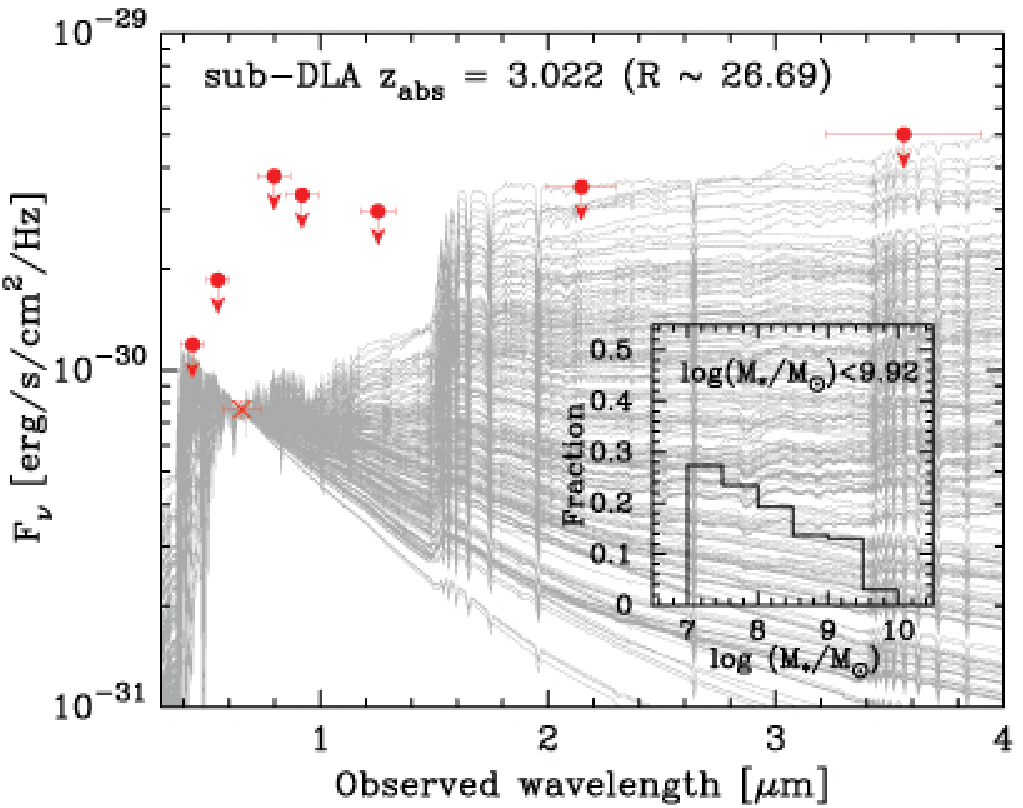}{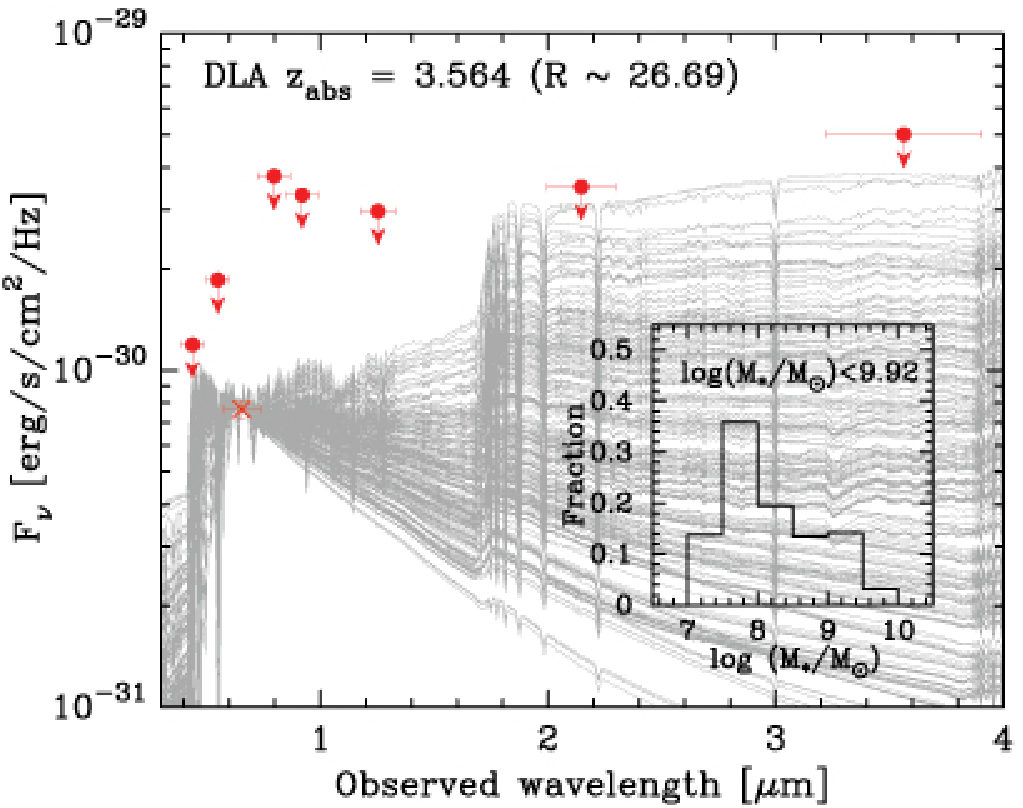}\\
\caption{Observed flux of G1 in the $R$ band (cross, $R\sim26.69$) and
  the $3\sigma$ flux upper limit in $BVIJK_sL$ bands (circles, see
  Table \ref{tbl-1}) are plotted as a function of the observed
  wavelength. We fit the model SEDs (solid lines) to the observed
  flux for the galaxies associated with the intervening DLA or \mgii\
  absorbers at $z\sim 1.773, 2.253, 3.022$, and  $3.564$. The model SEDs were
  drawn from the stellar population synthesis model of
  \citet{bc03} by considering a variety of star-formation histories
  (SFHs) including single-stellar populations, exponentially declining SFRs with $e$-folding timescales from 10 Myr to 5 Gyr, and
  constant star formation, and two different metallicities with $Z=
  Z_{\odot}$ and $0.02Z_{\odot}$. The models were computed for ages
  from 1 Myr to the cosmic age at each redshift. No dust reddening
  was assumed in the model SEDs. Fluxes of the model SEDs were scaled
  to fit the $R$-band flux of G1 (see the main text for details). Using
  the scaled flux and mass-luminosity ratio of the model SEDs, we
  constrained the stellar mass of the intervening DLA or \mgii\ host
  galaxies. The inset in each figure shows the distribution of the
  stellar masses derived from the model SEDs for each intervening
  absorber. 
  \label{mstar_rfix1}}
\end{figure*}%

\subsection{Spectroscopy of the Brightest Candidate}
To identify the redshift of the brightest object (G3), we performed
optical spectroscopy with FOCAS mounted on the Cassegrain focus of
the Subaru telescope \citep{kashik02}. We used a 0\farcs8 slit, 300B
prism, and SY47 order-sort filter, which provided a wavelength
coverage of 4800-9000 \AA\ with a spectral resolution of $\simeq$
9.6 \AA, to identify the rest-frame UV spectral features of the
absorbers at $z=2.0$-3.5, such as \lya\ emission, continuum
depression near the Lyman limit, and several interstellar absorption features seen in the spectra of LBGs \citep{shapley03}.
During target acquisition, we also took a $V$-band snapshot image
of the field around GRB 050730, including the three candidates, using
FOCAS (see Table \ref{tbl-1}). The total exposure times were 5400 s and 180 s for spectroscopy and snapshot imaging, respectively. Flux
calibration and atmospheric absorption-band correction were
performed using the spectrum of the spectrophotometric standard star
GD153, obtained on the same night. The spectroscopic data were reduced
in the standard manner using IRAF. We performed bias
subtraction, overscan subtraction, flat fielding with a dome lamp
flat, image distortion correction, wavelength calibration with a
ThAr lamp, background subtraction, and sky background
subtraction. Finally, the two-dimensional spectra were co-registered and combined,
and then the one-dimensional spectrum of the object was extracted from the combined
two-dimensional spectrum. In Figure \ref{phot-z}, we show the resultant FOCAS
spectrum of G3. We found no obvious rest-frame UV features in the obtained spectrum.

To further constrain the redshift of G3, we also estimated the
photometric redshift of the galaxy based on the photometry in $B$,
$V$, $R$, $I$,$z^{\prime}$, $J$, $K_s$, and $L$(3.6$\mu$m) bands
(points in Figure \ref{phot-z}) using the publicly available software
EAZY \citep{brammer08}. EAZY fits the observed SED of the galaxy with
a combination of galaxy templates. We used the default EAZY template
set eazy\_v1.0, which consists of five templates that span the colors
of galaxies in the semi-analytic model and an additional template to
compensate for the lack of young and dusty galaxies in the
semi-analytic model. Figure \ref{phot-z} shows the best-fit SED
template, which is in good agreement with both the photometric points
and the observed spectrum. The redshift of the best-fit SED is
$z_{\rm{phot}} = 0.519^{+0.042}_{-0.028}$. We also show the resulting
$\chi^2$ as a function of the photometric redshift, which reaches a
minimum at the redshift of the best-fit template ($z_{\rm{phot}} =
0.519$). Therefore, we conclude that G3 is most likely at
$z\sim0.5$ and is not related to the known intervening absorbers.

\section{CONSTRAINING THE PROPERTIES OF INTERVENING ABSORBERS} 
The above arguments suggest that G1 is most likely to be a host galaxy
of one of the intervening absorbers toward GRB 050730, whereas G2 and G3
are unlikely to be related to the absorbers. In this section, we
constrained the SFR and stellar mass for the four
intervening absorbers at $z=1.7$-3.5 toward GRB 050730 by assuming
that (1) G1 is a host galaxy of the absorber at each redshift or (2)
there is no detectable galaxy related to the absorbers. Finally, we
compared the constrained SFR and stellar mass of the intervening
absorbers at $z=1.7$-3.5 with the other high-z ($z>2$) galaxy
populations. 

\subsection{Star Fpormation Rate}
We estimated SFRs of the galaxies associated with the intervening DLA
or \mgii\ absorption systems at $z=1.773, 2.253, 3.022,$ and $3.564$
from the rest-frame UV continuum ($R$-band in the observed frame).  We
used the empirical formulae of \citet{madau98} assuming
\citet{salpeter55} IMF and no reddening. In Table \ref{tbl-2}, we show
the SFR for the absorbing galaxy at each redshift in case (1), the
case that G1 is a host galaxy of the absorbers. We also estimated the
SFR using the 3-$\sigma$ upper limit of the rest-frame UV continuum in
case (2), the case that there is no galaxy associated with the
absorbers. In all cases, we found that the SFR for the DLA galaxies at
$z>3$ was $\sim 2.5$ $M_{\odot}$ yr$^{-1}$ or smaller. At $z>2$,
\citet{peroux12} reported the upper limit of the SFR for typical less
metal-rich DLA host galaxies ($Z \leq 0.1 Z_{\odot}$), which are
similar in metallicity to our DLA ($z=3.564$) and sub-DLA ($z=3.022$)
samples, using the upper limit of the \ha\ fluxes derived from the
VLT/SINFONI IFU data. Our derived upper limit for less metal-rich DLA
host galaxies is comparable to, or even more stringent than, that
constrained by \citet[][$\sim$ 3.6 $M_{\odot}$ yr$^{-1}$ assuming
Salpeter IMF]{peroux12}.

In the case of $z\sim2$ \mgii\ system ($W_r^{\lambda2796} \sim 1$
\AA), we reached an SFR upper limit of 1.0 $M_{\odot}$ yr$^{-1}$, but
no host galaxies were clearly detected. By contrast, host galaxies of
strong \mgii\ systems ($W_r^{\lambda2796} > 0.3$ \AA) at $z\sim 1$
were detected with an SFR of more than 10 $M_{\odot}$ yr$^{-1}$
\citep{lovegrove11}. This difference in the SFR between $z\sim1$ and
$z\sim2$ was also argued in \citet{bouche12} for the case of a very
strong \mgii\ system with $W_r^{\lambda2796} \sim 2.0$ \AA. They
concluded that $z\sim2$ \mgii\ absorbers reside in smaller halos than
$z\sim1$ systems.  Although our sample size is only two, our result
for the strong \mgii\ systems with $W_r^{\lambda2796} \sim 1.0$ \AA\
supports the scenario proposed in \citet{bouche12}.

\subsection{Stellar Mass}
Using the GRB absorbers, we were able to strictly constrain the
continuum flux of the intervening absorbers, as the background GRB
afterglow used for identifying the absorbers had completely
disappeared when we performed the deep imaging. Thus, we did not need
to consider the ambiguity of the noise estimation arising from the
background PSF subtraction. Moreover, we could eliminate the possibility
that the galaxy associated with the absorbers was hidden in the
unsubtracted core of the background PSF, even if there was no
detectable galaxy around the position of the afterglow. These
unambiguous constraints on the continuum level provide a unique
opportunity to constrain the stellar mass of DLA or \mgii\ host galaxies
at $z>2$ for the first time.

The upper limits on the stellar masses associated with the intervening
DLA or \mgii\ absorption systems at $z=1.773, 2.253, 3.022,$ and
$3.564$ were derived by fitting the stellar population synthesis
models of \citet{bc03} to the upper limit for each object's continuum flux.
We used the same set of stellar population synthesis models as were used in Section \ref{color-color}. We derived the upper limit of the
stellar mass at each redshift for two separate cases: (1) we assumed that G1 was a host galaxy
of the absorber, and (2) we assumed that no galaxy was associated with the absorber.
Figure \ref{mstar_rfix1} shows the fitted model SEDs at each redshift
of the intervening absorption systems as a function of the observed
wavelength, assuming case (1). In the same figure, we also plot the
flux of G1 in the $R$ band and the 3-$\sigma$ upper limits on the fluxes
in the other observing bands. In case (1), the model SEDs were
scaled to fit the $R$-band flux of G1. In this case, we excluded any
model SEDs whose scaled fluxes exceeded any of the 3-$\sigma$ upper limits for the observed bands other than the $R$-band. In case (2), the model SEDs were scaled to avoid exceeding any of
the 3-$\sigma$ upper limits in the observed bands. Table \ref{tbl-2}
summarizes the results of the stellar mass estimations. We found that
the stellar mass of the DLA or \mgii\ absorbers at $z\gtrsim2$ is less
than a few times $10^9$ $M_{\odot}$, which is comparable to the mass
of dwarf galaxies like LMC \citep{harris09}.

It should be noted that our derived upper limits on the stellar mass
for the less metal-rich DLA at $z=3.564$ ([Si/H] $<-1.3$) and sub-DLA
at $z=3.022$ ([Si/H] $= -1.5\pm0.2$) are consistent with the stellar
masses predicted from the observed mass-metallicity relation for $z>2$
star-forming galaxies \citep[e.g.,][]{savaglio05, maiolino08}.

\begin{figure}
\epsscale{1.1}
\plotone{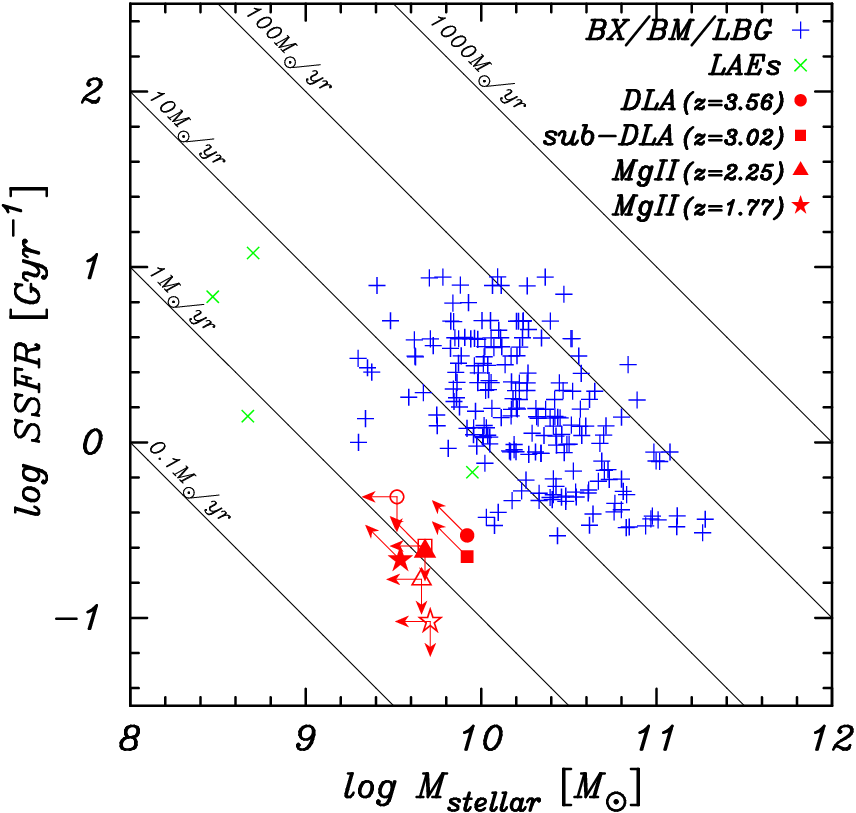}
\caption{Specific star formation rate (SSFR) as a function of stellar
  mass for DLA host galaxies at $z=3.022$ (squares), $3.564$ (circles) and
  \mgii\ host galaxies at $z=1.773$ (triangles), $2.253$ (stars). In the sample
  plot, we show the values for LBGs
  \citep[pluses]{reddy06,erb06} and LAEs \citep[crosses]{gawiser06,
    nilsson07, lai08} at $z=2$-$3$. Because the candidate of the host
  galaxy for each absorber is not clearly identified, we plot the
  SSFRs assuming that G1 is a host galaxy of each absorber
  (filled symbols) or that no galaxy in our image is associated with any
  absorber (open symbols). The solid lines show the SSFRs for
  constant SFRs from 0.1 to 1000 $M_{\odot}$ yr$^{-1}$. \label{ssfr}}
\end{figure}%

\subsection{Comparison with Other $z>2$ Galaxy Populations}
In this section, we compare the measured (or constrained) SFRs and
stellar masses for the DLA and \mgii\ host galaxies at $z_{\rm{abs}} \geq 2$
against those obtained for other $z>2$ galaxy populations, such as LBGs
\citep[e.g.,][]{reddy06,erb06} and LAEs \citep[e.g.,][] {gawiser06,
  nilsson07, lai08}. To compare the distribution of their SFRs and
stellar masses, we investigated their specific star formation rates
(SSFRs). The SSFR is defined as the ratio between the SFR and the stellar
mass, and it is an indicator of the intensity of star formation in the galaxy.
In Figure \ref{ssfr}, we plot the SSFR versus stellar mass for our DLA
absorbers at $z>3$ and \mgii\ absorbers at $z\sim2$. Given that we do not
identify a galaxy corresponding to each absorber, we constrained the
SSFR for each absorber in two separate cases: (1) we assumed that G1 was a host galaxy of the
absorber, and (2) we assumed that no host galaxy was detected in our image. For all cases, we found that our absorbing galaxies were located between the
low mass end of LBGs and LAEs. This implies that our sample absorbers,
which consist of less metal-rich ([Si/H] $< -1.0$) DLAs and \mgii\ systems, should not reside in the massive star-forming galaxies (like typical LBGs). Indeed, our derived SSFRs for the DLA/\mgii\ absorbers
are consistent with the SSFRs of $z=1.7-2.5$, less metal-rich DLAs, as
derived from their abundance pattern using star formation
histories (SFHs) similar to local irregular or dwarf starburst galaxies with
weak star formation efficiency \citep{dz07,calura09}.

\section{CONCLUSIONS}
We analyzed deep imaging data of the field around GRB 050730 in
optical to mid-infrared wavelengths to identify the host galaxies
giving rise to the intervening DLA and \mgii\ absorbers at $z=1.773,
2.253, 3.022$, and $3.564$. We used our own near-infrared and optical
imaging data obtained with the Subaru telescope, together with archived optical and mid-infrared imaging data obtained with the VLT
and $Spitzer$. We examined the color, impact parameter, and photometric
redshift (for the brightest galaxy only) for the galaxies detected in
our multi-color images to constrain the candidates for the galaxies
associated with the intervening absorbers. We found no
unambiguous candidate for the host galaxies in our data. Using the
$3\sigma$ limiting flux or the flux of the marginally detected galaxy
(G1) in the $R$-band image (corresponding to the UV wavelength in the
rest-frame of the intervening absorbers), we placed the following constraints on the upper limits of the SFR in the galaxies giving rise to the intervening absorbers:  $2.5$ $M_{\odot}$ yr$^{-1}$ for the DLA and sub-DLA systems and $1.0$ $M_{\odot}$ yr$^{-1}$ for the \mgii\ systems. We
used the 3-$\sigma$
upper limit of the flux in each wavelength to constrain the stellar mass of the absorbing galaxies to be
several times $10^9$ $M_{\odot}$ or lower. We confirmed that the
stellar mass limits for the DLA and sub-DLA at $z=3.564$ and
$3.022$ are consistent with the stellar masses expected from their
metallicity, assuming the mass-metallicity relation observed at
$z>3$ \citep{maiolino08}. Both the SFR and the stellar
mass for the intervening absorbers show properties similar to dwarf galaxies like the LMC \citep{harris09}. The intervening absorbers
in the direction of GRB 050730, which consist of the less metal-rich
DLAs and \mgii\ systems with $W_r^{\lambda2796} \sim 1.0$ \AA, show low SSFRs
($\lesssim 1.0$ Gyr$^{-1}$), which are comparable to LAEs or LBGs at
the low-mass end of their mass function. This low SSFR for our
intervening absorbers favors the scenario that typical DLAs and \mgii\
systems with low metallicities are likely to reside in dwarf galaxies,
rather than in massive star-forming galaxies like typical LBGs.

\acknowledgments 

We thank Kentaro Aoki and Takuji Tsujimoto for helpful discussions in
the early stage of this work and for careful reading of the manuscript.
This work is mainly based on data collected with the Subaru Telescope, which
is operated by National Astronomical Observatory of Japan
(NAOJ). Also, this work is based in part on archival data obtained
from the ESO Science Archive Facility and $Spitzer$ $Space$ $Telescope$,
which is operated by the Jet Propulsion Laboratory, California
Institute of Technology, under a contract with NASA.

\end{document}